# On the flow past an array of two-dimensional street canyons between slender buildings

Maria Grazia Badas[1], Simone Ferrari[1], Michela Garau[1], Alessandro Seoni[1], Giorgio Querzoli[1]

**Abstract** The flow above idealized, two-dimensional series of parallelepipedal buildings is examined with the aim of investigating how the building width ($W$) to height ($H$) aspect ratio affects the turbulence in the roughness sublayer and the ventilation of the underlying street canyons. We compare the case of buildings with a squared section ($AR_B = W/H = 1.0$) with a configuration with slender buildings ($AR_B = 0.1$) both in the case of unit canyon width ($D$) to height ($H$) aspect ratio ($AR_C = D/H = 1$) and in the case of $AR_C = 2$. The former corresponds to skimming flow and the latter to wake-interference regime. Measurements are performed in a water channel, measuring velocity on a vertical mid-plane using a particle image velocimetry technique. The mean flow, its second-order turbulence statistics, the exchange fluxes, and the integral time scales investigated, with results showing that slender buildings enhance turbulence production and yield longer integral time scales in the region just above the building roof. Namely, in the skimming-flow and wake-interference regimes, the maximum vertical velocity variance is more than doubled and increased by 50%, respectively. The combined analysis of the turbulence production fields and the snapshots of the flow during sweep and ejection events demonstrate that the shear layer between the canyon and the external flow is significantly more unstable with slender buildings, mainly because the damping effect of the vertical velocity fluctuations from the flat roof of the upwind building is substantially missing. Consequently, a larger (downstream) portion of the interface is prone to the direct interaction of the external flow structures. The higher turbulence intensity promotes the ventilation at the canyon interface, which is increased by a factor of two in the skimming-flow regime and a factor of 1.26 in the wake-interference regime. In summary, the present experiments show that the effect of the reduced building aspect ratio is particularly significant when the urban canopy consists of narrow canyons. The result is of interest since narrow street canyons are typically bounded by slender buildings in the urban texture of the old European city centres.

---

[1] DICAAR - Università di Cagliari - Via Marengo 2, 09123, Cagliari, Italy.

# 1 Introduction

Understanding the mechanisms that drive turbulent exchange between the urban canopy and the overlying boundary layer is crucial for assessing urban air quality and therefore the citizen exposure to pollutants released at the pedestrian level (Britter and Hanna 2003; Ng 2009; Fernando et al. 2010; Blocken et al. 2016). The complex urban morphology, resulting from the combination of the intricate street-network pattern and the heterogeneity of building geometry, is known to considerably affect the turbulent flow which, in turn, determines the pollutant dispersion (Fernando et al. 2001; Kastner-Klein et al. 2004). Thus, an accurate estimation of flow characteristics in an urban environment should consider the real geometry of the investigated area (Carpentieri and Robins 2015). However, both the experimental and numerical flow simulations in the case of realistic geometries suffer from the drawbacks of being expensive and giving information that cannot be easily extended to other configurations (Zajic et al. 2015).

Therefore, in order to gather information that can be generalized to large sets of configurations, many studies have been performed on simplified geometries, focusing on the effect of particular morphological features on the turbulent flow and dispersion. Initially, the effect of the street-canyon aspect ratio both in two- and three-dimensional configurations has been investigated (Oke 1988; Hagishima et al. 2009; Zajic et al. 2010; Di Bernardino et al. 2014, 2015; Simón-Moral et al. 2014). Other researchers described the importance of the building shape, including details about the building morphology (Richardson and Surry 1991; Kanda and Maruta 1993; Ozmen et al. 2016; Ferrari et al. 2017; Garau et al. 2019). Recently, Murena and Mele (2016), with their numerical simulations, found that the presence of balconies can determine a significant modification in the flow field inside two-dimensional street canyons, which, in turn, produces less effective mass exchange with the atmosphere above. Llaguno-Munitxa et al. (2017) confirmed the negative role of facade elements on the turbulent mixing above a building array, whereas they observed that complex roof geometries tend to increase the mixing. Numerical simulations described by Badas et al. (2017), and water-channel experiments of Garau et al. (2018) highlighted how, in presence of two-dimensional canopy, pitched roofs modify the turbulence structure inside the canyons and deepens the shear layer at their upper interface.

The building aspect ratio (i.e., the ratio of the width to the height of the building), has received much less attention, and most of the above-mentioned configurations were based on squared section buildings. However, as the slope of the roof has a decisive effect in redirecting the flow just upstream from the canyon and changing the interfacial shear-layer characteristics, so the length of a flat roof (non-dimensionally coinciding with the building aspect ratio) may change the airflow and the turbulent exchange. Actually, Hosker (1984) described the flow patterns near isolated buildings of different aspect ratios subject to a perpendicular incident wind, showing how the presence of a roof recirculating region and its possible reattachment depends on the building aspect ratio. In the case of an urban canopy of rectangular-prism-shaped obstacles, Sadique et al. (2017) examined the influence of roughness-element aspect ratio on the mean and turbulent velocity fields and pointed out how its effect is different in the case of aligned and staggered obstacles.

Boundary-layer modifications due to regular wall roughness, aligned in the spanwise direction, also have received much attention in the field of turbulent channel flows. However, most research was focused on the effect of varying the fluid gap between the solid bars, i.e., the canyon aspect ratio (Cui et al. 2004; Leonardi et al. 2007). Macdonald et al. (2018) performed direct numerical simulations (DNS) of high aspect ratio spanwise-aligned bars in a turbulent boundary layer. They showed that for deep cavities, when the obstacle height is greater than three times the bar spacing, the overlying flow only depends on the bar spacing and not on the bar height. This would imply that the building aspect ratio has an influence on the above flow up to a threshold value. These results were obtained from DNS in the case of d-type roughness (corresponding to skimming flow), and considering obstacles with equal width and spacing. Other authors studied the effect of the upstream turbulence by varying the geometry of the roughness upwind of the investigated two-dimensional canyon (Perret et al. 2017; Blackman et al. 2018). Jaroslawski et al. (2019) deepened the same issue, focusing on the spanwise structure of the flow and its interaction with large-scale structures, which is fundamental to predict lateral flows. However, the above studies considered only roughness elements consisting of cubes or squared section bars. In the case of a squared section canyon, the interaction between the cavity flow and the overlaying boundary layer was observed to be mediated by a shear layer that develops horizontally from the vertex of the upstream wall (Perret and Savory 2013). Salizzoni et al. (2011) analyzed the shear-layer instability, concluding that it depends both on the velocity difference across the shear layer and the length scales of the turbulence.

However, to the best of the authors' knowledge, the influence of the building aspect ratio was not specifically addressed in the case of a two-dimensional urban canopy. Cai et al. (2008) compared two-dimensional urban canopy simulations both consisting of squared and rectangular section buildings (specifically, the building width was one third of the building height). Comparing the results of these simulations, the authors stated that the effect of building width is insignificant on the flow with respect to canyon width. Nonetheless, this statement was used only to justify the adoption of a reduced computational domain and this conclusion cannot be generalized. Moreover, the possible influence of building aspect ratio can be interesting, especially in densely populated urban areas, such as the historical centre of European cities, where narrow and long street canyons are often bounded, on their sides, by tall and thin buildings. An example of the heterogeneity of urban morphometric parameters can be found in Badas et al. (2019), who performed an analysis of an Italian town; the resulting distributions of planar and frontal area indexes, which are strictly linked to both canyon and building aspect ratio, show how the most studied squared section building departs from real data. Hence, it would be extremely important, also for practical purposes, to assess whether and how the flow pattern and street ventilation are affected by the building aspect ratio.

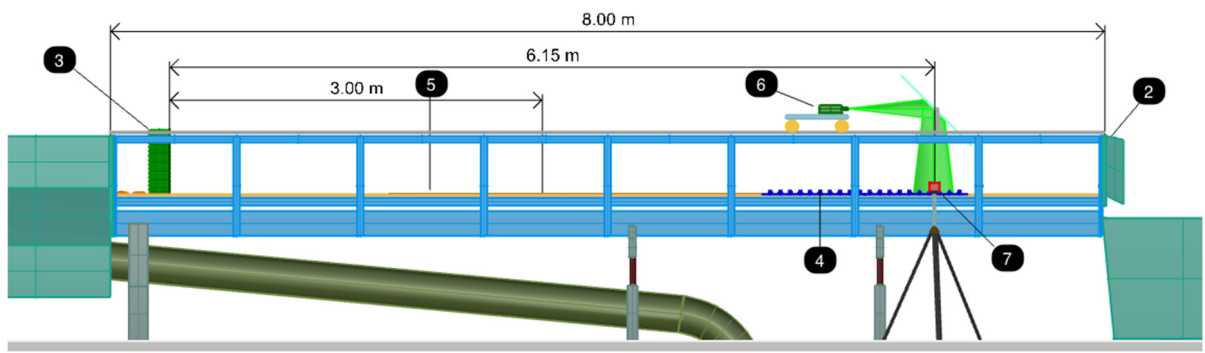

**Fig. 1** Water channel at the Hydraulics laboratory of the University of Cagliari. (1) constant head reservoir; (2) sharp-crested weir; (3) honeycomb grid; (4) prismatic buildings; (5) pebbles; (6) laser; (7) measurement region

In the present study, we experimentally investigate the flow past arrays of two-dimensional street canyons with the aim of elucidating the role of the building aspect ratio on the airflows within the urban canopy and the overlaying boundary layer. We compare the typical squared section prismatic buildings with the flow over canyons bounded by very slender buildings, i.e., whose width is one tenth of their height. We describe the effects of the building aspect ratio in terms of mean velocity field, turbulence statistics, and airflow exchanges.

## 2 Experimental Set-up and Procedures

Experiments were carried out in the recirculating water channel of the Hydraulics Laboratory at the University of Cagliari sketched in Fig. 1. The experimental facility and measuring techniques are the same as those used by Garau et al. (2018). However, for the sake of clarity, we recall here the main characteristics of the set-up and procedures. The channel (8.00-m long, 0.40-m wide, and 0.50-m high) is fed by a constant head reservoir (1); the flow rate was regulated by a sharp-crested weir (2) placed at the end section of the channel; at the initial section, a honeycomb grid (3) eliminated any secondary flows. The measurement area was located 6.15 m downstream of the honeycomb, where an array of 20 prismatic buildings (4) spanning over the whole channel width, modelled the idealized two-dimensional urban canopy. Small pebbles (5-mm average size) were irregularly glued on the portion of the channel bottom between the honeycomb and the building array (5) in order to promote the rapid development of a turbulent boundary layer. Two street-canyon aspect ratios were considered: the first, $AR_C = 1$, corresponding to the flow regime described by Oke (1988) as skimming flow, and the second, $AR_C = 2$, characterized by the regime of wake interference, where $AR_C = D/H$, $D$ is the canyon width, and $H$ is the building height (see Fig. 2).

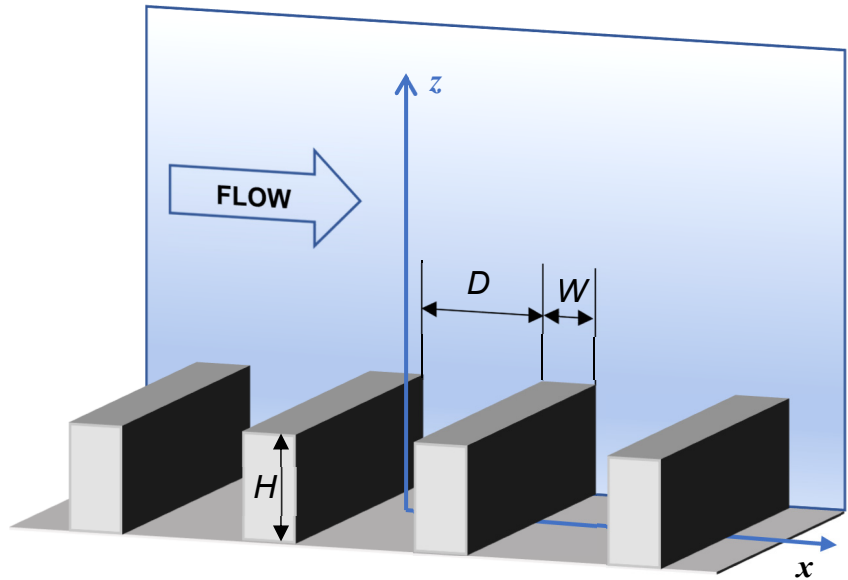

**Fig. 2** Sketch of the two-dimensional urban canopy and reference frame. Axis origin is located at ground level in the centre of the investigated canyon

For each canyon aspect ratio, two building shapes were considered. The first consisted of 20 × 20 mm$^2$ squared-section prisms; therefore the building aspect ratio, $AR_B$, was equal to one, i.e., the shape extensively considered in literature (Mo and Liu 2018; Di Bernardino et al. 2018). The building aspect ratio, $AR_B$, is here defined as the ratio of the building width, $W$, to the height, $H$ (see Fig. 2). The second series of prismatic obstacles had a 2 × 20-mm$^2$ section resulting in an aspect ratio $AR_B = 0.1$, corresponding to very slender buildings. The measurement area was centred on the 17$^{th}$ canyon, allowing the boundary layer to reach its equilibrium over the 16 canyons upstream of the test section (Brown et al. 2000; Llaguno-Munitxa et al. 2017; Garau et al. 2018). The characteristics of the flow upstream of the obstacles are reported in Garau et al. 2018. However, since in the measurement area the boundary layer is fully developed, the characteristics of the incoming flow coincide with those measured at the test canyon itself. All results are described in the reference frame drawn in Fig. 2, with $x$ axis orientated streamwise, the $z$ axis vertically upwards. Correspondingly, in the following, $u$ and $w$ indicates respectively the $x$ and $z$ velocity components.

The two velocity components, $u$ and $w$, were measured using a non-intrusive image-analysis technique. The vertical midsection of the channel was illuminated by a light sheet (2-mm thick) generated by a 532-nm diode laser, 2 W in power (6). A high-speed camera (1726 × 2240 pixel in resolution), placed orthogonally to the measurement plane (7), recorded 310 grey-scale images per second. The region framed by the camera was 0.10-m wide and 0.08-m high, resulting in a spatial resolution of 40 μm per pixel. The water was uniformly seeded with pine pollen particles with an average diameter of 20 μm, which are neutrally buoyant in water (Miozzi and Querzoli 1996). The velocity field was obtained by means of a feature tracking algorithm that analyzes pairs of successive frames in three main steps: firstly, the particle positions are identified on the first frame by means of the Harris algorithm (Harris and Stephens 1988). Secondly, the tracking is performed by comparing squared interrogation windows centred on the particles' locations to shifted windows on the successive frame and looking for the shift minimizing the dissimilarity between windows. The dissimilarity was measured using an algorithm based on the Lorentzian estimator, averaged over the interrogation windows (in the present experiments, the window size was 19 × 19 pixels). Thirdly, the samples were validated by an algorithm based on a Gaussian filtering of first neighbours, defined by the Delaunay triangulation. The method is robust to high spatial velocity gradients and appearance/disappearance of particles. A detailed description of the dissimilarity measurement algorithm, with an evaluation of the performance compared to the classical cross-correlation used in the PIV technique, is reported by Falchi et al. (2006), whereas the tracking procedure is described by Besalduch et al. (2013, 2014).

For each canopy configuration we acquired a total of 48,000 frames at 310 Hz framerate, thus yielding 48,000 instantaneous velocity fields. With the aim of increasing the statistical robustness of the dataset, we divided the acquisition in $N = 40$ recordings of 1200 frames (i.e., lasting 3.9 s each) separated by a time interval long enough for the acquisitions to be statistically independent. Statistics on a regular grid (0.06 $H$ × 0.06 $H$) were computed by considering separately all the instantaneous scattered velocity samples that fall on each grid cell. A statistical convergence test was performed by computing the mean and the variance of the streamwise velocity component in the cell centred in $x = 0.0\ H$ and $z = 1.0\ H$ by using an increasing number, $n$, of 1200-frame recordings, and plotting the relative deviation from the value obtained with the whole dataset (i.e., $n = N$),

$$\frac{|\bar{u}(n)-\bar{u}(N)|}{\bar{u}(N)}; \frac{|\overline{u'^2}(n)-\overline{u'^2}(N)|}{\overline{u'^2}(N)}.$$

We measured an average of 830 velocities, within the considered grid-cell area, per recording dataset. Consequently, the total number of samples (within the cell) exceeded 33500. Results, presented in Fig. 3, show that the relative deviation remains stably below 3%, provided that $n$ exceeds 30 recordings, thus confirming the statistical robustness of the dataset.

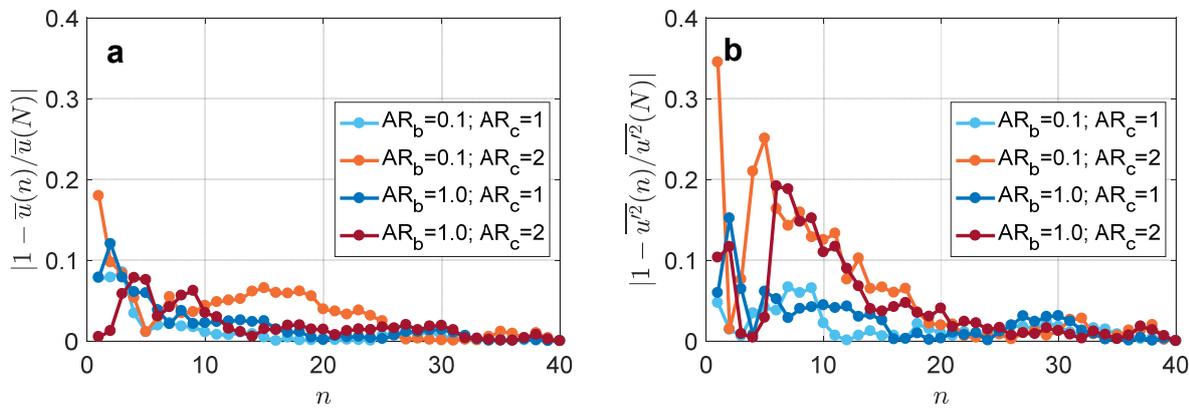

**Fig. 3** Relative deviation of the mean (panel a) and variance (panel b) of the streamwise velocity component in the grid cell centred in $x = 0.0\ H$, $z = 1.0\ H$ as a function of the number, $n$, of 1200-frames recordings included in the dataset

During the present experiments, the water in the investigation area was 0.42-m deep and the freestream velocity magnitude was $U = 0.35$ m s$^{-1}$. The Reynolds number based on the freestream velocity magnitude and the building height, $H$, resulted: $Re = UH/\nu = 7,000$. Previous tests, performed on the flow approaching the building series, revealed that, following the argument proposed by Uehara et al. (2003), only a 1.0-mm-deep layer adjacent to the wall is affected by a significant decrease of the velocity because of the viscosity effects. Therefore, the bulk of the flow can be assumed Reynolds independent (Garau et al. 2018).

## 3 Results
### 3.1 Mean Velocity

Figure 4 shows the overall structure of the mean velocity field in the four investigated geometrical configurations in terms of non-dimensional out-of-plane vorticity, $\omega H/U$ (colour maps), and streamlines. The y-component of the vorticity, $\omega$, was calculated by means of a centred difference scheme applied to the mean velocity field on the regular grid described in the previous section. Irrespective of the building aspect ratio, when $AR_C = 1$ (Figs. 4a and 4c), the mean field shows the typical features of the skimming-flow regime (Oke 1988), with the region between the buildings dominated by a single clockwise vortex centred in the canyon. Similarly, in both the cases with $AR_C = 2$, the flow pattern exhibits the typical characteristics of the same flow regime, i.e., the wake-interference regime, which consists of a large clockwise vortex on the downwind side of the cavity and a smaller counter-clockwise vortex at the lower upwind corner of the canyon. The above observations confirm, as expected, that the aspect ratio of the prismatic buildings does not affect significantly the mean flow regime. However, at a closer look, some differences in the vorticity distribution can be identified.

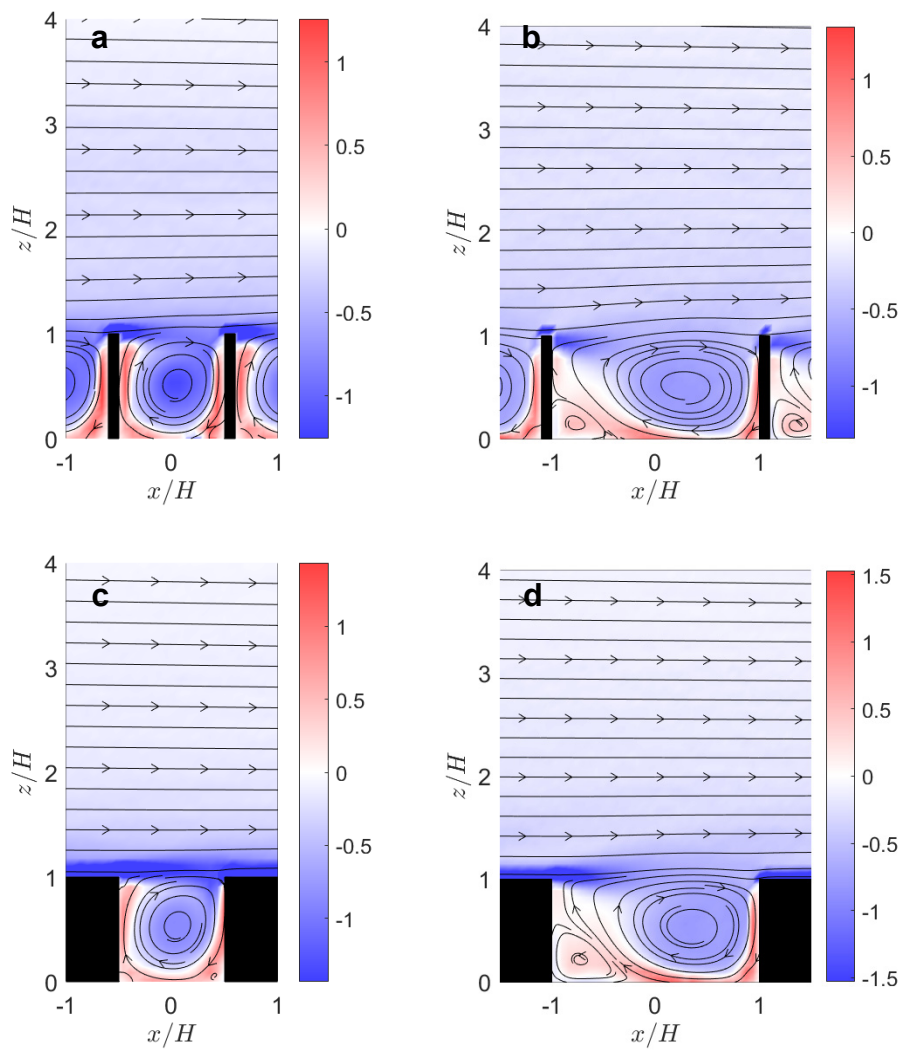

**Fig. 4** Streamlines of the mean velocity field and non-dimensional out-of-plane vorticity, $\omega H/U$, for the four investigated configurations. Panel a: $AR_B = 0.1$, $AR_C = 1$; panel b: $AR_B = 0.1$, $AR_C = 2$; panel c: $AR_B = 1.0$, $AR_C = 1$; panel d: $AR_B = 1.0$, $AR_C = 2$

In the skimming-flow regime, the square section buildings ($AR_B = 1.0$) generate a continuous layer of high negative vorticity along the whole interface between the canyon and the overlaying flow. In that region, the high vorticity corresponds to an intense shear layer. Differently, in the case of slender buildings ($AR_B = 0.1$), the vorticity layer decreases rapidly from the roof of the upwind building, and it is almost undistinguishable in the downwind portion of the canyon interface, as a consequence of a less sharp shear layer along most of the interface. The comparison of the flow within the cavity shows higher vorticity levels both at the building walls and in the centre of the main vortex compared to the case $AR_B = 1.0$, indicating that the recirculation within the canyon is more intense when it is bounded by slender buildings. That stronger recirculation is an indication of a more efficient momentum transfer from the overlaying flow to the cavity, which in turn, following the argument by Bentham and Britter (2003), can be considered a clue of a higher exchange of air between the canyons and the outer flow, i.e., a better ventilation. The hypothesis will be confirmed by the further analysis presented below.

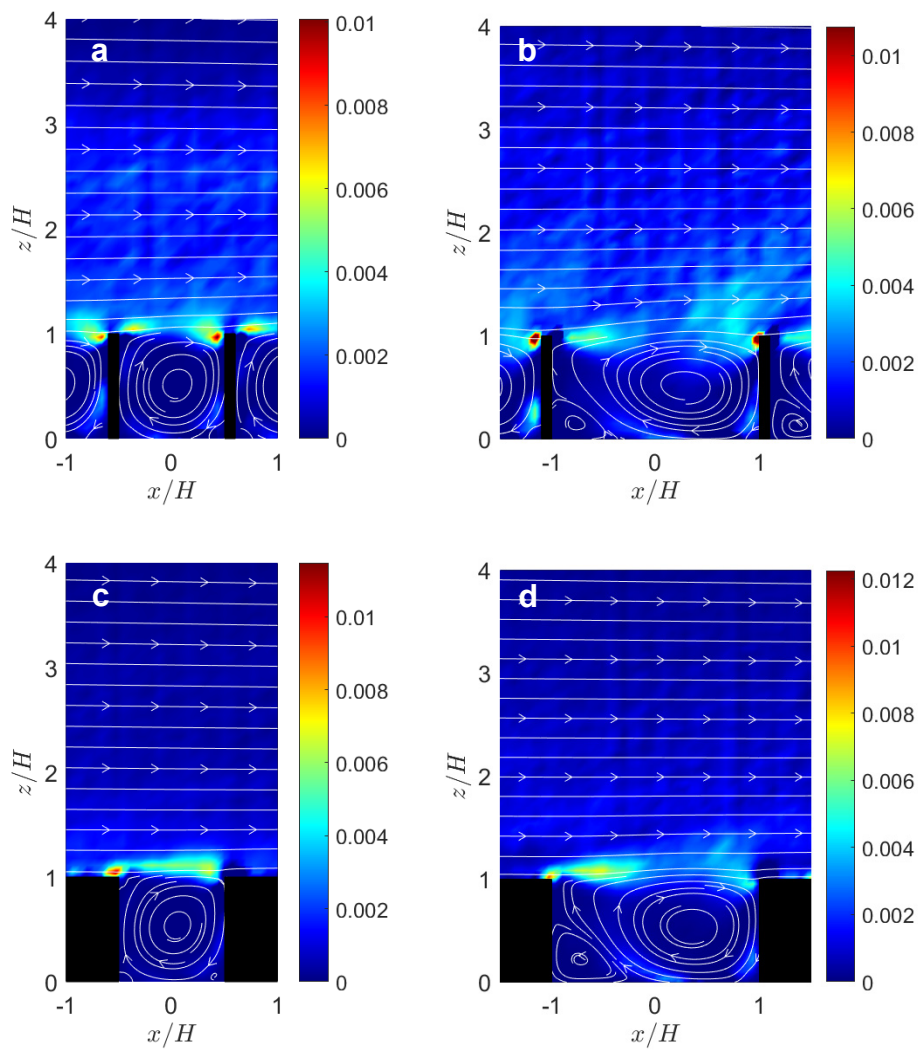

**Fig. 5** Non-dimensional production of turbulence kinetic energy, $P$. Streamlines are superimposed as white lines. Panel a: $AR_B = 0.1$, $AR_C = 1$; panel b: $AR_B = 0.1$, $AR_C = 2$; panel c: $AR_B = 1.0$, $AR_C = 1$; panel d: $AR_B = 1.0$, $AR_C = 2$

The differences are less pronounced in the wake-interference regime. At that canyon aspect ratio ($AR_C = 2$), the vorticity at the upper canyon interface is less intense irrespective of the building aspect ratio, suggesting a less sharp shear layer separating the cavity flow from the overlaying boundary layer. Consequently, the vorticity layer propagating from the roof of the upwind building vanishes before the mid-width of the canyon. The intensity of the vorticity layer has a slightly higher intensity in the case of squared section buildings. Conversely, the intensity of the vorticity within the canyon is almost independent of the building aspect ratio.

**3.2 Turbulence**

In order to elucidate how the building shape affects the generation of turbulence, the local value of the non-dimensional production term of the turbulence kinetic energy budget was computed and mapped in Fig. 5. Since the measurement plane coincides with the symmetry mid-plane of the channel, the non-dimensional production term, $P$, reduces to:

$$P = -\left[\overline{u'^2}\frac{\partial \overline{u}}{\partial x} + \overline{u'w'}\left(\frac{\partial \overline{u}}{\partial z} + \frac{\partial \overline{w}}{\partial x}\right) + \overline{w'^2}\frac{\partial \overline{w}}{\partial z}\right]\frac{H}{U^3} \quad (1)$$

where $u$ and $w$ indicate respectively the velocity components along the $x$ and $z$ axis (orientated as defined in Fig. 2) and, as usual, the overbar and primes indicate respectively the Reynolds average and the fluctuations around the average. All the statistics are computed for each cell of the grid described in Sect. 2. The kinetic energy production, $P$, has been made non-dimensional by $U$, and the building height, $H$. The same expression holds under the assumption of virtually infinite length of the canyon in the spanwise direction.

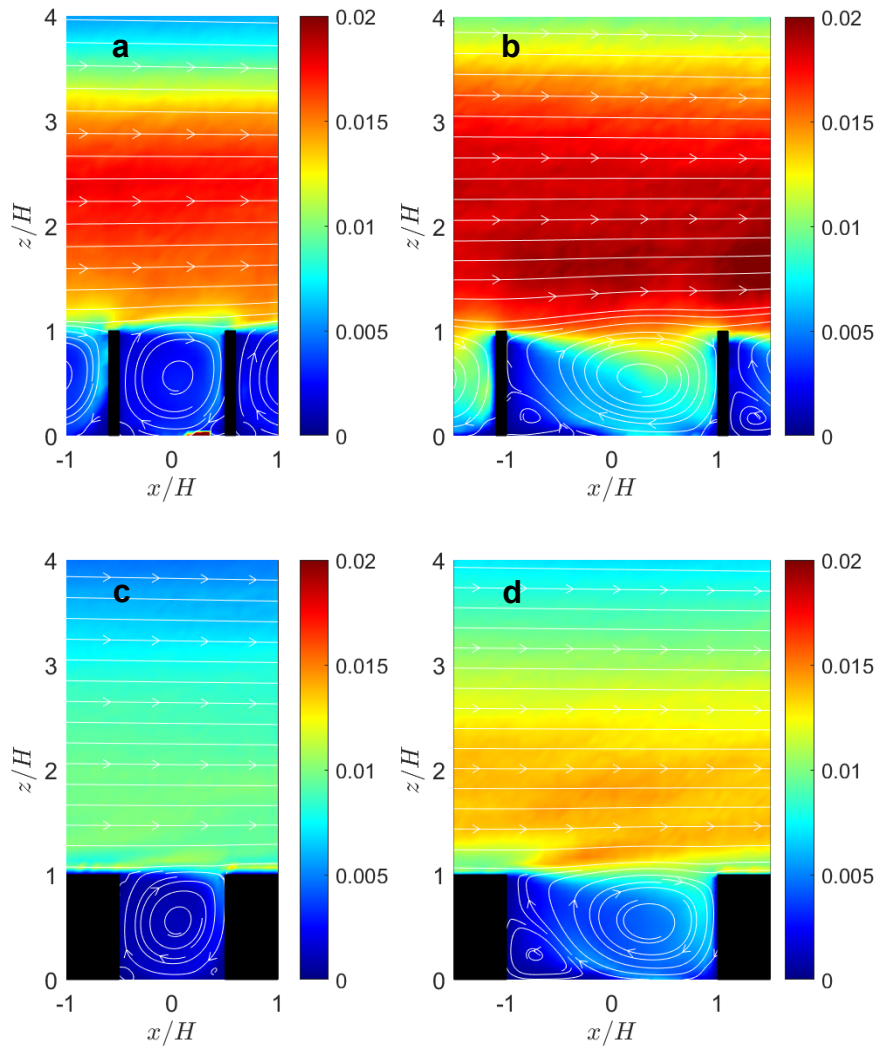

**Fig. 6** Maps of turbulence kinetic energy, $E$, made non-dimensional by the square of freestream velocity, $U$. Streamlines are superimposed as white lines. Panel a: $AR_B = 0.1$, $AR_C = 1$; panel b: $AR_B = 0.1$, $AR_C = 2$; panel c: $AR_B = 1.0$, $AR_C = 1$; panel d: $AR_B = 1.0$, $AR_C = 2$

As expected, most of turbulence production occurs at the shear layer which develops at the roof level. In further detail, the region of turbulence production propagates, while increasing in depth, from the upper corner of the upstream building. The production is weaker in the central zone of the interface, where the underlying mean flow is mainly horizontal, and increases in correspondence to the zones where the vertical velocity component dominates the cavity flow. With squared section buildings, the turbulence production has a small vertical extension, with some vertical spread only at its downstream end, particularly in wake-interference regime. Conversely with the slender buildings, in addition to the interfacial layer, the turbulence production extends over the region above the canopy, reaching significant intensities, especially just upstream of the windward façade.

The changes in the distribution of the production result in a quite different distribution of turbulence in the roughness sublayer. As we performed planar velocity measurements, Fig. 6 shows the two-dimensional turbulence kinetic energy obtained from the in-plane velocity components, made non-dimensional by the square of freestream velocity magnitude,

$$E = \frac{1}{2U^2}\left(\overline{u'^2} + \overline{w'^2}\right). \qquad (2)$$

Previous measurements pointed out that, with squared section buildings, the turbulence in the roughness sublayer (i.e., the region where the flow is influenced by the individual roughness elements) is more intense in the case of wake interference compared to skimming flow (Garau et al. 2018). Colour maps shown in Fig. 6 demonstrate that the decrease of the building aspect ratio down to $AR_B = 0.1$ does not alter this scenario. However, irrespective of the canyon aspect ratio, the flow above the canopy is meaningfully more turbulent when the canopy consists of slender buildings. In the skimming-flow regime the turbulence is always most confined out of the cavity. Differently, in the wake-interference regime the presence of slender buildings promotes the penetration of the turbulence over the entire region above the diagonal from the top of the leeward wall to the base of the windward wall; whereas, in the case of square buildings, the high turbulence zone extends only in the region just in front of the windward façade.

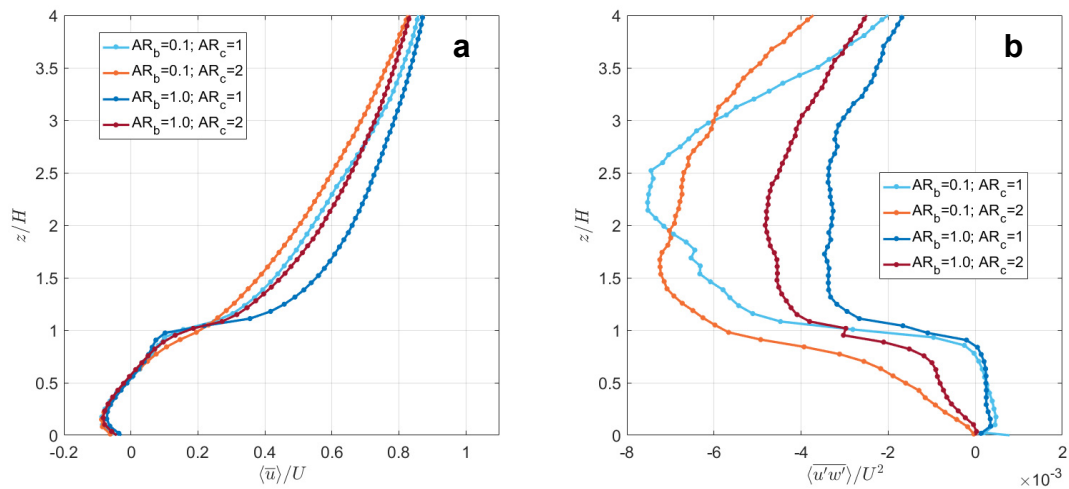

**Fig. 7** Vertical profiles of horizontally-averaged mean streamwise velocity (panel a) and momentum flux (panel b) for the four configurations displayed in the legend. The plots are made non-dimensional by $U$ and $U^2$, respectively

### 3.3 Vertical Profiles

The maps presented above describe the phenomenon in all its spatial details. However, it is often useful to consider the canopy as a whole, leaving out the effects of the individual roughness elements. For example, most of the urban canopy models that are available today assume spatially averaged velocity profiles (Macdonald et al. 1998; Kim et al. 2015; Coceal and Belcher 2016). Following that approach, in this section we present the horizontal averages of the main quantities describing the mean flow and turbulence in the roughness sublayer.

Following the canopy approach (Finnigan 2000), we computed the horizontal average over a periodic canopy unit,

$$\langle \cdot \rangle = \frac{1}{\lambda(z)} \int_{-\lambda/2}^{\lambda/2} (\cdot) dx, \qquad (3)$$

with $\lambda(z) = D + W(z)$ for $z > H$ and $\lambda(z) = D$ for $z \leq H$ (see Fig. 2). Therefore, we considered the whole periodic-unit length above the buildings and limited the averaging to the canyon width below the building height.

The vertical profiles of the spatially-averaged streamwise velocity component (Fig. 7a), $\langle \bar{u} \rangle$, collapse on a single curve, irrespective of the aspect ratio of the canyon and the building. Some differences can be observed for $0.8H < z < H$, i.e., at the shear layer, corresponding to the differences in the intensity of the shear layer observed also in the vorticity maps (Fig. 4). The slope of the profiles in that region confirms that the velocity gradient generated by the skimming-flow regime ($AR_C = 1$) is higher compared to the gradient generated by the wake-interference regime ($AR_C = 2$). However, at a given canyon aspect ratio, the interfacial velocity gradient is higher when delimited by squared section buildings ($AR_B = 1.0$).

Above the building height the mean velocity exhibits different levels of velocity defect that, not surprisingly, are directly correlated to the values of the negative turbulent momentum flux attained in the region just above the canopy (Fig. 7b). In fact, wake interference gives highest values of momentum flux, and higher velocity defect, compared to skimming flow (see e.g. Di Bernardino et al., 2015) and the presence of slender buildings enhances momentum fluxes above the canopy compared to squared section buildings, at a given canyon aspect ratio. Additionally, Fig. 7b, shows that within the squared section canyon the turbulent momentum flux is slightly negative at the top of the canyon cavity and positive elsewhere inside the canyon, irrespective of the building shape, in agreement with the findings of Di Bernardino et al. (2015) and works referenced therein. Differently, in the case of wake-interference regime, $\langle \overline{u'w'} \rangle$ is negative and monotonically increasing throughout the whole cavity depth. In the wake-interference regime, the slender buildings give meaningfully higher flux values compared to squared section buildings.

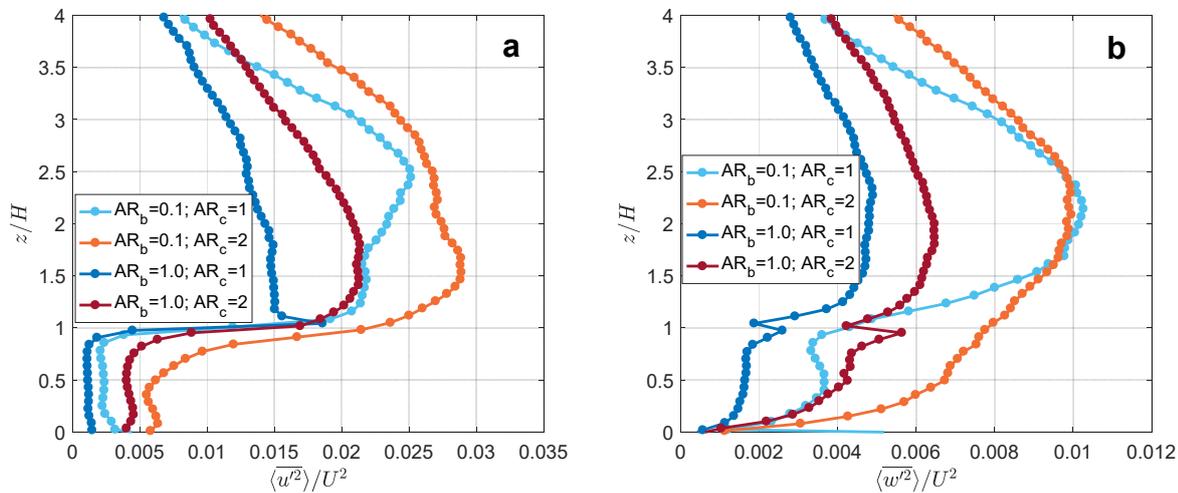

**Fig. 8** Non-dimensional variance of the streamwise (panel a) and vertical (panel b) velocity components for the four configurations displayed in the legend

In agreement with what is observed in the maps of turbulence kinetic energy shown in Fig. 6, the streamwise and vertical velocity variance profiles also are significantly influenced by the building aspect ratio. Some features are common to the variance of both the components: i) along the whole height of the roughness sublayer, the slender buildings tend to increase the variance for both flow regimes; ii) within the canyon, the effect of the flow regime prevails on that of the building shape, and the variance of the cases corresponding to skimming flow are lower than those in wake-interference regime in accordance with other experiments performed with squared-section buildings (Kastner-Klein et al. 2001); iii) conversely, above the building height, the building shape prevails, and the variance with slender buildings is significantly higher than in the case of squared section buildings, irrespective of the canyon aspect ratio.

In the skimming-flow regime, the streamwise velocity variance is a maximum at $z = H$ for squared section buildings, whereas the maximum is shifted up to about $z = 2.5H$ with slender buildings. Differently, in the wake-interference regime the maximum, although increased in value, remains at $z = 1.5H$ when $AR_B$ is decreased from 1.0 down to 0.1. Within the canyon all the configurations show a similar behaviour.

The vertical velocity variance is a maximum at about $2H$ with $AR_B = 1.0$, while the maximum is located at $z = 2.2H$ with $AR_B = 0.1$. In the wake-interference regime, the region of high vertical-velocity variance protrudes more deeply within the canyon when the buildings are slender, indicating that the propagation of $E$ within the cavity observed in Fig. 6b is mainly due to vertical velocity fluctuations.

Figure 9 displays the vertical profiles of the correlation coefficient,

$$r_{uw} = \frac{\overline{u'w'}}{\sqrt{\overline{u'^2}}\sqrt{\overline{w'^2}}},\qquad(4)$$

horizontally averaged over the urban periodic unit, which can be interpreted as a measure of momentum transport efficiency. Similarly to turbulence kinetic energy, within the canyon, the profiles are mainly dependent on flow regime, whereas above the building height the values are similar but the influence of building shape prevails. Above the canopy top, $r_{uw}$ is nearly constant for all the examined cases, and both the cases with slender buildings present higher values, witnessing more momentum transport per unit variance compared to the other conditions. In the cavity, in the presence of skimming flow, $r_{uw}$ presents positive values except near the cavity top, whereas in wake-interference regime $r_{uw}$ follows the same monotonical trend toward the axis origin. This trend resembles the family portrait of forest canopies studied by Böhm et al. (2013).

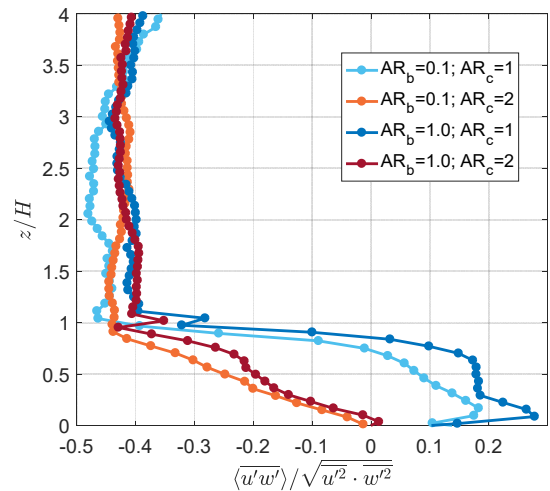

**Fig. 9** Correlation coefficient spatially averaged on the canyon periodic unit for the four configurations indicated in the legend

### 3.4 Exchange Flux

The vertical air exchanges within the street canyon, and especially at its top, are crucial for the urban air quality since the capability of the flow to remove the pollutant released at the street level, for instance because of the vehicular emissions, basically depends on them. There are different ways to estimate the exchange flux. All of them are based on integrating the velocity field over a given surface (usually the upper canyon interface) in order to compute the air flux. Some authors take into account only the mean velocity field (see e.g. Neophytou et al. 2014), while some others consider only the velocity fluctuations, under the assumption that the mean flow, due to the mass conservation, does not contribute to the air exchange (Liu et al. 2005). We followed the approach based on the integration of the instantaneous velocity proposed by Weitbrecht et al. (2008) in the context of river flows, and used also to study cavity flows (Tuna et al. 2013) and street canyons (Perret et al. 2017).

Let $\lambda(z)$ be the horizontal section of the canyon at height $z$. We compute the vertical air exchange flux, $\varphi_e(z)$, as:

$$\varphi_e(z) = \frac{1}{2} \int_{\lambda(z)} |w(t)| dx, \qquad (5)$$

i.e., the time average of the positive and negative vertical fluxes that, though instantaneously different, are equal on average (Garau et al. 2018). In addition, we compute the contribution of the sole mean velocity field to the exchange air flux as:

$$\varphi_{em}(z) = \frac{1}{2} \int_{\lambda(z)} |\overline{w}| dx. \qquad (6)$$

The vertical profiles of $\varphi_e$ and $\varphi_{em}$, made non-dimensional by the canyon width, $D$, and the freestream velocity, $U$, are presented in Fig. 10. The turbulent contribution may be inferred as the difference between $\varphi_e$ and $\varphi_{em}$. The overall behaviour of the exchange flux is basically insensitive to the building shape, predominantly depending on the flow regime. In the SF regime, $\varphi_e$ and $\varphi_{em}$ are at a maximum at about mid canyon-height, where the vertical velocity dominates, and decrease both towards the ground and the roof level. Conversely, the contribution of the turbulence attains its higher values in the shear layer at the interface between the canyon and the overlaying boundary-layer flow. Neophytou et al. (2014) computed the exchange flux based on mean velocity field in a similar experimental arrangement, considering two-dimensional canyons between squared-section buildings at $0.5 < AR_C < 2.33$. The values are not quantitatively comparable, as far as they chose the bulk velocity through the flume as the velocity scale. However, they found values at roof height about doubled in the case of $AR_C = 2$ compared to the case of $AR_C = 1$, whereas in our experiments $\varphi_{em}$ is about the same. The difference can be explained by the higher velocity defect of the horizontally averaged velocity profiles in the case of wake interference with respect to skimming flow (see Fig. 7a) which, in turn, corresponds to a decrease of the bulk-to-freestream velocity ratio.

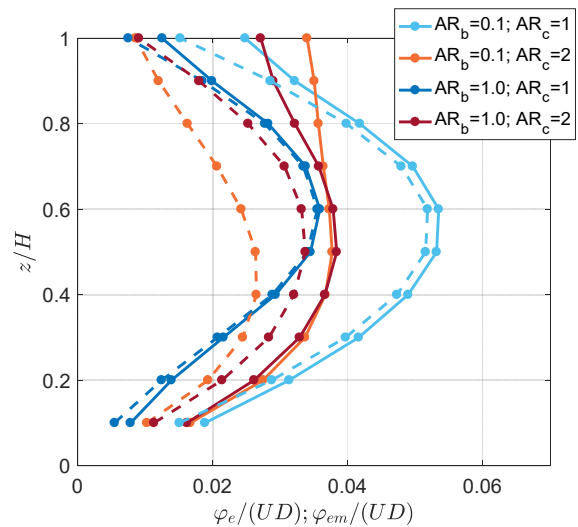

**Fig. 10** Non-dimensional vertical profiles of exchange fluxes, $\varphi_e$ (solid lines), and $\varphi_{em}$ (dashed lines), made non-dimensional by the freestream velocity, $U$, and the canyon width, $D$. Line colours indicate the geometry of the roughness as reported in the legend

As above mentioned, the higher velocity defect in Fig. 7a results from a more efficient vertical momentum transfer which is known to be related to an increased exchange flow rate (Bentham and Britter 2003). Thus, it is not surprising that the increase measured by Neophytou et al. roughly corresponds to the one we observed in terms of total non-dimensional exchange flux $\varphi_e/(UD)$. Perret et al. (2017) measured the exchange flux at the rooftop height in different canyon and upstream roughness configurations, including $AR_B = AR_C = 1.0$. In that configuration they found $\varphi_e/(UD) = 0.021$, quite higher compared to our measurements ($\varphi_e/(UD) = 0.013$). However, the turbulence in the canopy sublayer is known to depend on the characteristics of the upstream flow also (Blackman et al. 2018). A comprehensive summary of studies on flow exchange rate is reported in Kubilay et al. (2017). Though it should be stressed that they are obtained with different methods, most of values reported are in reasonable agreement with the present results.

In Sect. 3.1 we observed that the vertical transfer of momentum towards the canyon was more efficient in the presence of slender buildings, in particular for square-section canyons, and we conjectured that the air exchange would be as well. The plot in Fig. 10 confirms the speculation. Though the overall behaviour seems to be insensitive to the building shape, the vertical fluid exchange is enhanced at all heights in the presence of slender buildings in the skimming-flow regime, and mainly in the upper part of the canyon for the wake-interference regime. Comparison of $\varphi_e$ (solid lines) and $\varphi_{em}$ (dashed lines) shows that in all the analyzed configurations the contribution of the turbulence is increased when $AR_B = 0.1$. However, the increase of the contribution of the turbulence to the ventilation is particularly high for the wake-interference flow regime, thus compensating for the decrease of $\varphi_{em}$ observed for $AR_B = 0.1$ and $AR_C = 2$.

### 3.5 Integral Time Scales

In order to describe the time scale of the coherent structures dominating the roughness sublayer, we computed the spatial distribution of the 1/e time scales of the horizontal ($T_u$) and vertical ($T_w$) velocity over the periodic roughness unit. In every point of a regular grid ($\Delta x = 0.04H$, $\Delta z = 0.1H$), we computed the time auto-correlation coefficients of the two measured velocity components and estimated the 1/e time scale as the time lag when the correlation coefficient decreases to 1/e, which corresponds to the integral time scale, provided that the autocorrelation decreases exponentially. Maps of horizontal, $T_u$, and vertical, $T_w$, 1/e scale are presented in Figs. 11−14, made dimensionless by the advective time $H/U$.

The spatial distribution of $T_u$ within the canyon (Figs. 11 and 12) is linked to the topology of the mean flow, thus, not particularly influenced by the building aspect ratio. Not surprisingly, regions of high values are found: i) where the flow is dominated by horizontal, slow motions; ii) at the centre of the main recirculating vortices; iii) at the interface between two adjacent vortices. However, excluding these regions, the 1/e scale within the canyon is lower compared to the region above the roof height. The presence of slender buildings tends to shorten the time scale, irrespective of the canyon aspect ratio. In general, the upper interface of the canyon is characterized by short time scales, indicating that the region is dominated by the small vortical structures generated by the shear layer. Conversely, in the case of wake-interference regime and $AR_B = 0.1$ (Fig. 11b), relatively high values extend from the external flow to the centre of the main vortex, indicating that the region is dominated by the entrainment of external turbulent structures that, while propagating through the cavity, tend to preserve their time scale.

Similar considerations can be drawn for the vertical velocity time scale, $T_w$ (Figs. 13 and 14): the spatial distribution varies according to the topology of the flow, which in turn is mainly affected by the canyon aspect ratio. The three points illustrated above hold with the only difference that high values of $T_w$ are found in the regions dominated by slow, vertical flow. It should be also noticed that the external flow exhibits always shorter vertical time scales, $T_w$, compared to the flow within the canyon.

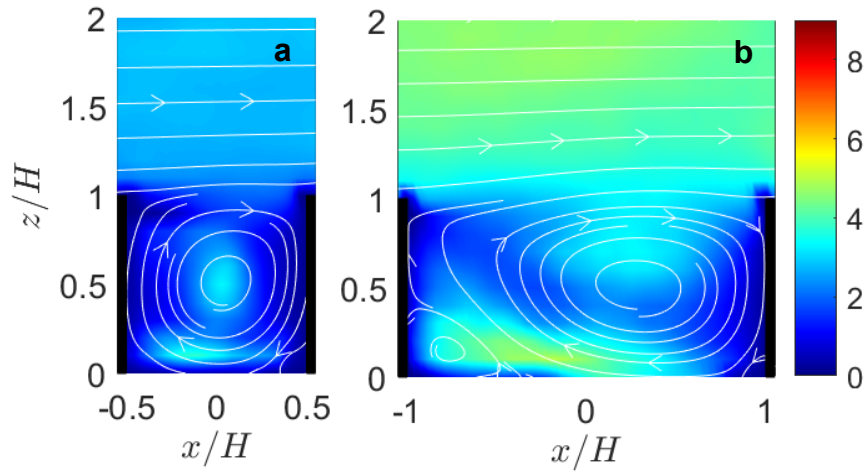

**Fig. 11** Map of the dimensionless 1/e time scale of the horizontal velocity, $T_u U/H$, for the building aspect ratio $AR_B = 0.1$. Panel a: $AR_C = 1$; panel b: $AR_C = 2$

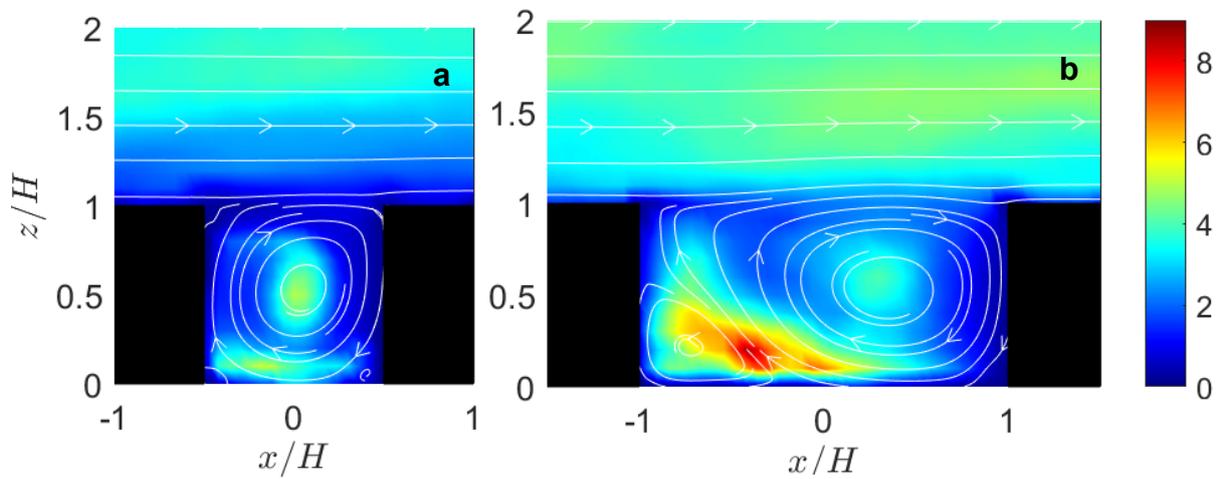

**Fig. 12** Map of the dimensionless 1/e time scale of the horizontal velocity, $T_u U/H$, for the building aspect ratio $AR_B = 1.0$. Panel a: $AR_C = 1$; panel b: $AR_C = 2$

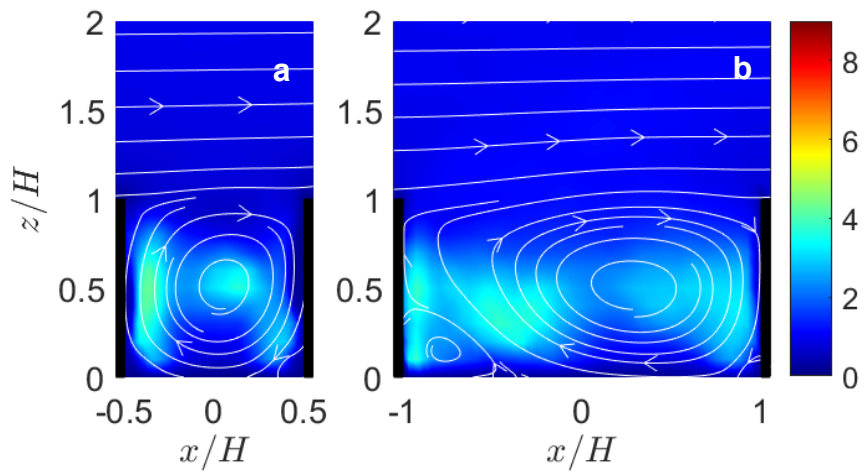

**Fig. 13** Map of the dimensionless 1/e time scale of the vertical velocity, $T_w U/H$, for the building aspect ratio $AR_B = 0.1$. Panel a: $AR_C = 1$; panel b: $AR_C = 2$

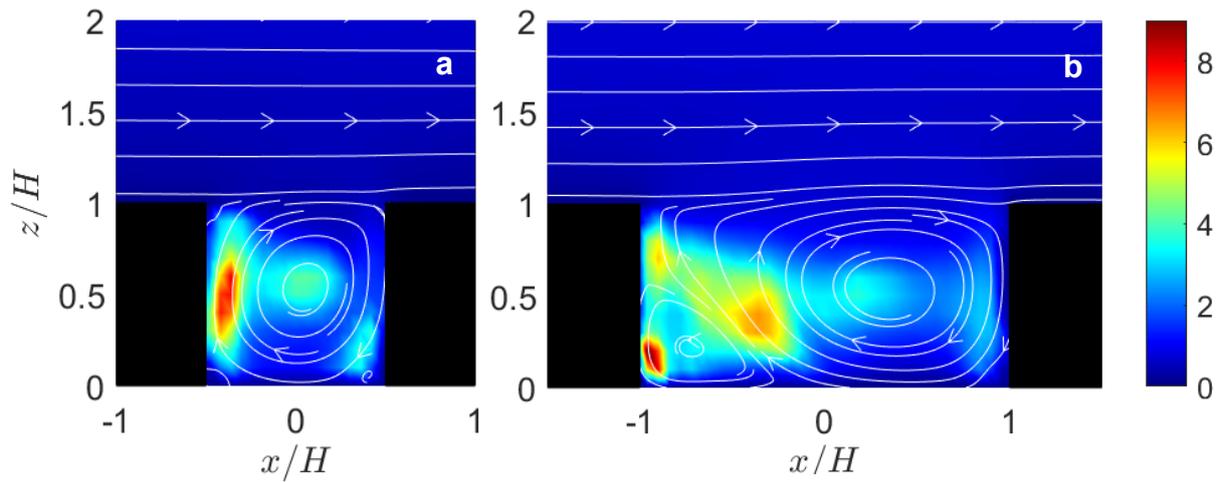

**Fig. 14** Map of the dimensionless 1/e time scale of the vertical velocity, $T_w U/H$, for the building aspect ratio $AR_B = 1.0$. Panel a: $AR_C = 1$; panel b: $AR_C = 2$

Even in this case the presence of slender buildings has the effect of decreasing the peak values of 1/e time scale within the canyon both in the skimming-flow and wake-interference regime.

Figure 15 shows the vertical profiles of the non-dimensional 1/e time scales averaged horizontally over a periodic roughness unit. The analysis of the profiles confirms the main features yet observed in Figs. 11−14. $<T_u>$ is a maximum at $z = 0.1H$ irrespective of the aspect ratios. Maximum values are observed in the wake-interference regime. However, the peak value is significantly decreased in the presence of slender buildings. The case of SF and $AR_B = 1.0$ displays the minimum values at the shear layer ($z \approx H$), indicating the presence of a sharp shear layer characterized by small structures that separates the cavity flow from the overlaying boundary layer in agreement with the findings of Blackman et al. (2015). In general, the slender buildings tend to increase the horizontal velocity time scale at $z = H$, suggesting a shear layer dominated by larger vortical structures, which are supposed to promote the mixing between the canyon and the external flow.

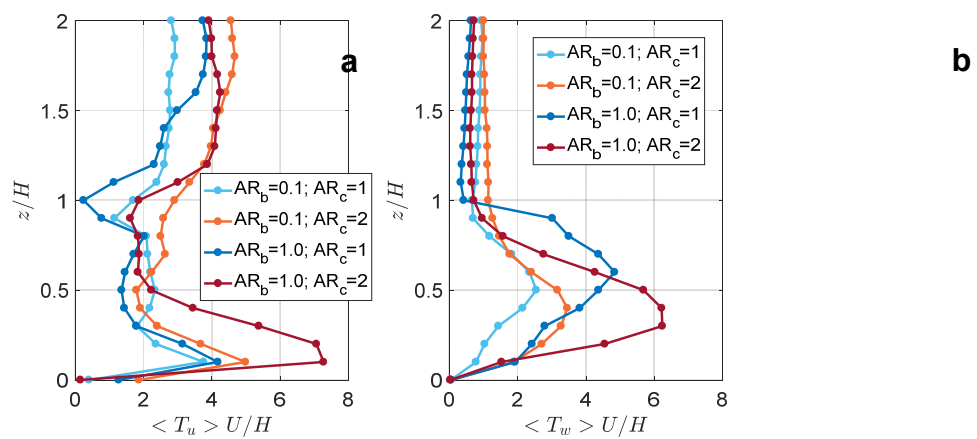

**Fig. 15** Non-dimensional profiles of horizontally-averaged 1/e time scales. Panel a: scale of streamwise velocity component, $T_u$; panel b: scale of vertical velocity component, $T_w$. Time scales are made non-dimensional by the advective time $U/H$

For both the building aspect ratios, above $z \approx 1.4H$ the time scale is mainly controlled by the shape of the canyon, and the wake-interference regime gives longer time scales of the horizontal velocity compared to skimming flow (see also Table 1, reporting the values of the horizontal and vertical velocity time scale at $z = 2.0H$). The result is in agreement with the findings of Salizzoni et al. (2011) and, more recently, Blackman et al. (2018), which measured the integral length scales above an array of two-dimensional, squared-section buildings ($AR_B = 1.0$), and observed that the turbulence is characterized by smaller scales in the case of skimming flow compared to the wake-interference regime.

**Table 1** Values of non-dimensional, horizontally averaged 1/e time scale of the horizontal, $<T_u>$, and vertical, $<T_w>$, velocity component at $z = 2.0H$

| $AR_B$ | $AR_C$ | $<T_u>H/U$ | $<T_w>H/U$ |
|---|---|---|---|
| 0.1 | 1 | 2.8 | 0.9 |
| 0.1 | 2 | 4.5 | 1.0 |
| 1.0 | 1 | 3.7 | 0.6 |
| 1.0 | 2 | 3.9 | 0.7 |

The time scale of the vertical velocity component, $T_w$, is diminished within the canyon by the presence of slender buildings for both the flow regimes (Fig. 15b). In that case, the effect of the building shape prevails on that of the canyon aspect ratio. Therefore, the peaks attained with $AR_B = 0.1$ are lower compared to $AR_B = 1.0$ regardless of the flow regime. In the roughness sublayer, at $z = 2.0H$, the value of $T_w$ is mainly driven by the building shape (Table 1) and the time scale tends nearly to the same value irrespective of the canyon aspect ratio. The slender buildings generate the longest time scales.

## 4 Discussion

Results show that when the street canyon is delimited by slender buildings, significantly higher production of turbulence kinetic energy is observed above the canopy (Fig. 5) giving rise to a higher level of turbulence in that region (Fig. 6). Correspondingly, the shear layer separating the flow within the canyon from the overlaying roughness sublayer tends to be smoother, mainly in the downwind portion of the interface (Figs. 4 and 7a), and it is characterized by longer time scales (Fig. 15). The higher turbulence intensity corresponds to an increased vertical flux of horizontal momentum (Fig. 7b).

In the skimming-flow regime ($AR_C = 1$), the more effective momentum transfer from the external flow to the vortex dominating the cavity in the case of slender buildings is witnessed also by the higher vorticity characterizing the cavity flow (Figs. 4a and 4c).

In the wake-interference flow regime, the presence of slender buildings promotes the entrainment of the turbulent structures from the roughness sublayer deep into the upper-downwind half of the canyon (Figs. 6b and 11b), in particular because of the increase of the vertical velocity fluctuations at the canyon interface and below (Fig. 8b).

**Table 2** Non-dimensional air exchange flux at the canyon interface ($z = 1.0$)

| $AR_B$ | $AR_C$ | $\varphi_e/(UD)$ |
|---|---|---|
| 0.1 | 1 | 0.025 |
| 0.1 | 2 | 0.034 |
| 1.0 | 1 | 0.013 |
| 1.0 | 2 | 0.027 |

The higher turbulence intensity and momentum flux significantly improve the ventilation at the intermediate canyon heights in the case of SF. Namely, the maximum $\varphi_e$ within the canyon is increased 1.5 times in that case (Fig. 10), thus promoting the rapid dispersion of the pollutant released in the proximity of the street level. Conversely, in the case of WI regime, the increase of the turbulent contribution to the ventilation just compensates for the corresponding decrease of the contribution of the mean flow. Finally, slender buildings cause a significant increase of the exchange flux, $\varphi_e$, at the interface ($z = H$) irrespective of the flow regime, thus improving the overall ventilation of the canyon with external fresh air (Table 2).

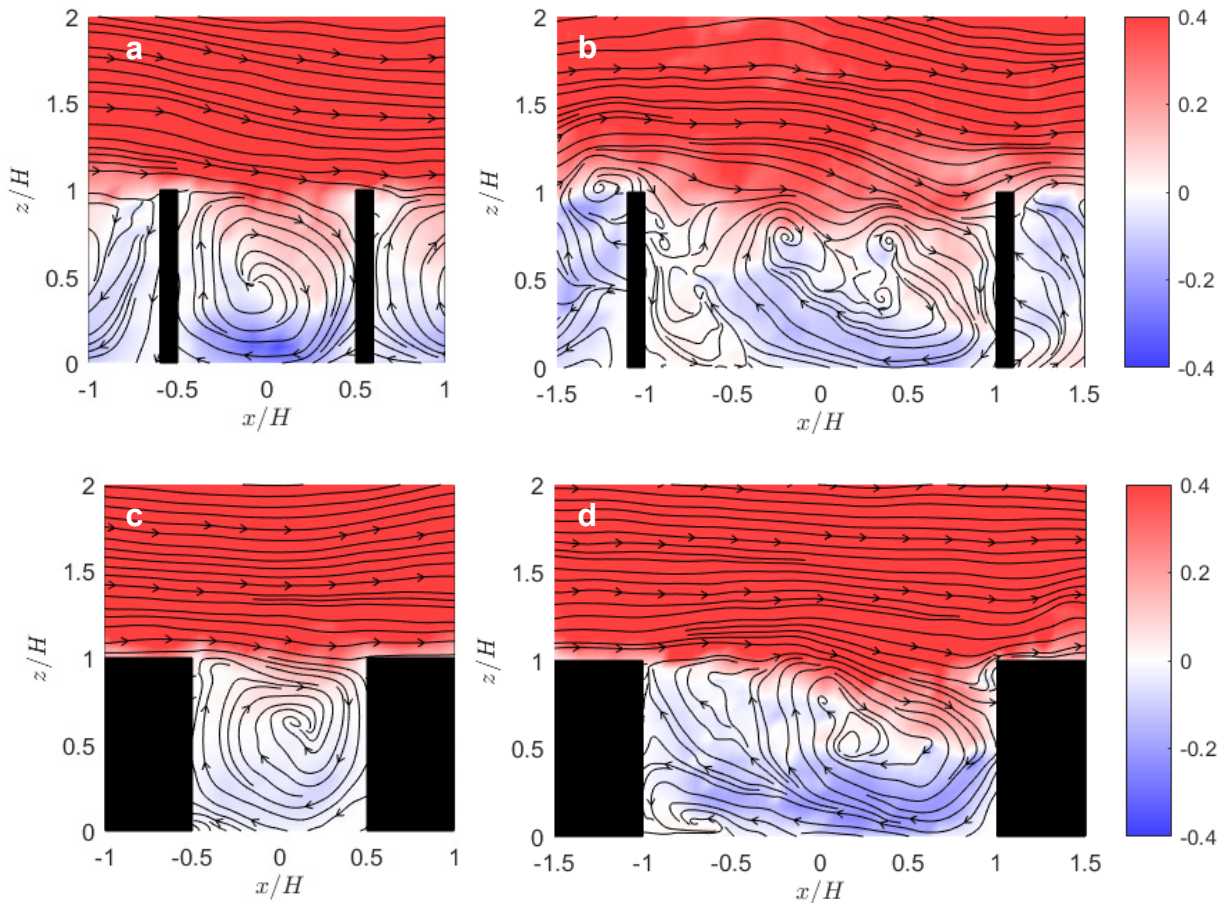

**Fig. 16** Instantaneous streamlines during typical sweep events. Colours indicates the non-dimensional streamwise velocity. Panel a: $AR_B = 0.1$, $AR_C = 1$; panel b: $AR_B = 0.1$, $AR_C = 2$; panel c: $AR_B = 1.0$, $AR_C = 1$; panel d: $AR_B = 1.0$, $AR_C = 2$

Salizzoni et al. (2009) investigated the mechanism driving the ventilation in a squared section canyon with two-dimensional roughness elements with different spacings upstream. They observed that a shear layer develops over the entire upper, canyon interface, which develops Kelvin−Helmholtz instabilities and interacts with the recirculating flow within the cavity and the coherent structures dominating the overlaying boundary layer. Blackman et al. (2018), in similar conditions, found evidence of non-linear interaction between the inclined large-scale structures of high or low momentum detected in the boundary layer and the small scales close to the canopy. This scenario is confirmed by the present results. The region of high values of the production of turbulence kinetic energy, $P$, shown at $z = H$ with the squared section cavity and $AR_B = 1$ (Fig. 5c), clearly indicates the presence of the above-mentioned shear layer, which develops from the upwind building vertex through the whole interface length, while getting unstable and increasing in depth. However, for the same cavity geometry but with slender buildings (Fig. 5a), the high production region developing from the upwind vertex diffuses rapidly and most of the interface is not characterized by a clearly recognizable shear layer. The different development of the interface flow suggests a different mixing mechanism.

In order to elucidate the phenomena driving the above-mentioned change, we plotted the instantaneous flow fields during typical sweep (Fig. 16) and ejection (Fig. 17) events. Inspection of Fig. 16c and 17c shows that, in the case of squared-section buildings, sweep and ejection events are due to the flapping of the interfacial shear layer, which however preserves its coherence. In that case, it is reasonable to assume that the mixing at the interface is governed by the shear-layer instability, possibly triggered by the large-scale structures of the external boundary layer (Blackman et al. 2018) With slender buildings the sweep mechanism does not change (Fig. 16a); conversely, the ejection (Fig. 17a) cannot be ascribed to a flapping motion of the interface but it is the consequence of the upward motion along the upwind cavity wall. In that case, the mixing is driven by the dynamics of the recirculating flow within the cavity, which in turn is triggered by the downwards-momentum injections at the downwind cavity wall which occur during the sweep events.

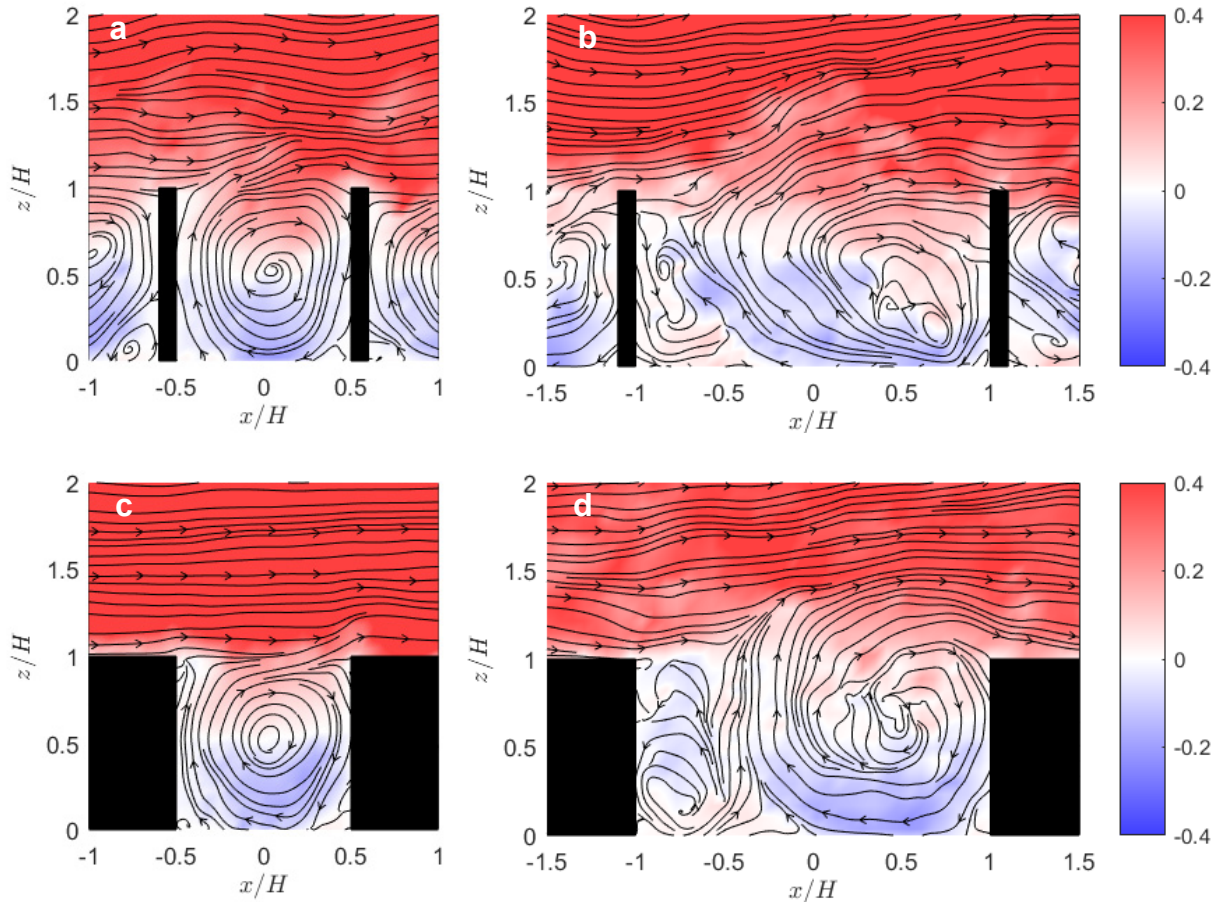

**Fig. 17** Instantaneous streamlines during typical ejection events. Colours indicate the non-dimensional streamwise velocity. Panel a: $AR_B = 0.1$, $AR_C = 1$; panel b: $AR_B = 0.1$, $AR_C = 2$; panel c: $AR_B = 1.0$, $AR_C = 1$; panel d: $AR_B = 1.0$, $AR_C = 2$

The reason for the different behaviour is that, with squared-section obstacles, the horizontal, flat roof of the upstream building (which is $1.0H$ long) inhibits the vertical velocity fluctuations of the fluid generating the upper portion of the shear layer, which thus is subject to fewer perturbations. However, this effect is almost completely missing with slender buildings since the roof length is only $0.1\ H$. Conversely, the vertical perturbations originating from the upstream canyon can propagate towards the developing shear layer without obstacles.

The comparative analysis of the turbulence production maps of Fig. 5 and the instantaneous flow fields of Figs. 16 and 17 helps to understand the dynamics of the interface also in the case of WI.

With $AR_C = 2$ and squared section buildings, Fig. 5d shows a shear layer that starts from the upwind vertex and develops downwind for a length of about $H$. Further downstream, the turbulent production is weaker and spreads up to $1.4H$ and no evidence of the presence of a well-defined shear layer is found. The instantaneous flow fields show that, during both the sweeps (Fig. 16d) and ejections (Fig. 17d), the flow within the cavity is basically separated from the external flow by the shear layer from $x = -1.0H$ to $x = -0.1H$ (where mixing is governed by the shear layer dynamics). Conversely, in the downwind half of the interface, the external flow enters the cavity directly during the sweep events (triggered by the large-scale structures of the boundary layer), whereas ejections are driven by the vertical upwards motion of the large recirculating vortex. For slender buildings the scenario is not different but for the reduced length of the shear layer in the upwind portion of the interface (due to the missing of the vertical-velocity fluctuation inhibition by the flat roof, Figs. 6b, 16b and 17b). As a consequence, the portion of the interface where the mixing is not mediated by the shear layer, and the interaction between external and cavity flow is direct, is larger, thus producing a higher exchange flow.

## 5 Conclusions

With the purpose of investigating the effect of the building aspect ratio on the airflow above an urban canopy, we performed a series of experiments measuring the velocity field in the flow past canopies of different shapes. We focused on the case of long street canyons separated by slender buildings, which are typical of the densely populated urban areas. Since our aim was to gather general indications about the phenomenon, we considered idealized, periodical, two-dimensional canopies. We compared the classical squared-section two-dimensional roughness elements ($AR_B = 1.0$), to the limit case of reduced aspect ratio buildings ($AR_B = 0.1$), both in the skimming-flow ($AR_C = 1$) and wake-interference ($AR_C = 2$) flow regime.

Results show that both in the skimming-flow and wake-interference regime, the presence of slender buildings enhances the turbulence in the overlaying boundary layer and the canyon ventilation. The effect is more evident in the case of the skimming-flow regime. In particular, the analysis of the vertical profiles of the spatially averaged quantities shows that the maximum horizontal velocity variance is increased by a factor of 1.7 in skimming-flow regime and by a factor of 1.4 in the wake-interference regime. Similarly, the maximum vertical velocity variance is increased by a factor of 2.1 with skimming flow and by a factor of 1.5 in the wake-interference regime.

The combined analysis of the production of the turbulence kinetic energy, $P$, and some snapshots of the flow field taken during sweep and ejection events, revealed that, with slender buildings, the interfacial shear layer is more unstable, thus promoting a higher exchange flux at the rooftop level.

The increase given by the presence of slender building is by a factor 2.0 in the regime of skimming flow and a factor 1.3 in the regime of wake interference. The slender buildings yield also a significantly higher vertical fluid exchange within the canyon in the case of skimming flow (the maximum value is increased by a factor 1.5), whereas, in the wake-interference regime, the overall exchange flow remains basically unvaried as far as the significant increase in the turbulence contribution provided by the slender buildings just compensates for the decrease of the contribution from the mean motion. In conclusion, though a parametric study would be required to explore quantitatively the change in the ventilation as a function of the building aspect ratio, the present experiments indicate that the role of the reduced building aspect ratio is crucial in the case of narrow street canyons. This result is particularly interesting since narrow street canyons are typically delimited by slender buildings in the urban texture, for example, of the old European city centres.

**Acknowledgements** This research has been funded by the Autonomous Region of Sardinia, FSC 2014-2020, grant number RASSR50082.